\documentclass[conference]{IEEEtran}
\ifCLASSINFOpdf
\else
\fi

\usepackage{algorithmic}

\usepackage{epsfig}
\usepackage{xspace}
\usepackage{times}

\usepackage{bibspacing}
\begin{document}
\title{Towards a General Software Engineering Methodology for the Internet of Things}

\author{
\IEEEauthorblockN{Franco Zambonelli}
\IEEEauthorblockA{
	Dipartimento di Scienze e Metodi dell'Ingegneria\\
	Universit\'a di Modena e Reggio Emilia\\
	Reggio Emilia, Italia\\
	franco.zambonelli@unimore.it }
}

\maketitle

\begin{abstract}
As research in the Internet of Thing area progresses, and a multitude of proposals exist to solve a variety of problems, the need for a general principled software engineering approach for the systematic development of IoT systems and applications arises. In this paper, by synthesizing form the state of the art in the area, we attempt at framing the key concepts and abstractions that revolve around the design and development of IoT systems and applications, and draft a software engineering methodology centered on these abstractions. 
\end{abstract}

\section{Introduction}
\label{introduction}

The possibility of enriching physical objects and places with wirelessly accessible sensing, computing, and actuating capability enables the ''Internet of Things'' (IoT) vision \cite{Atzori10,gubbi2013}. This defines a scenario in which everything in our physical and social worlds can become the node of a large-scale situated network, supporting coordinated actions to sense and control the physical world itself and to facilitate our interactions with it \cite{Con12}.   

The dramatic future impact of IoT in society, industry, and commerce is already widely recognized \cite{Por14,Ian14}. However, despite the great deal of worldwide researches in the area, the technologies to make IoT a systematic reality are far form being assessed. Early researchers in the IoT area have mostly focussed on communication issues and on enabling interoperability \cite{Atzori10}. More recently, a great deal of effort has been devoted at promoting sound approaches to facilitate the integration of resources and services towards the provisioning of software defined distributed services for the IoT. For instance, as in the so called ``Web of Things'' vision \cite{wottuto,wot15}, by promoting the provisioning of resources in an IoT network in term of Web Services, and thus making it possible to develop distributed and coordinated IoT services by using standard Web technologies. 

WoT approaches are definitely promising will very likely represent a keystone technologies in the future of IoT. Indeed, along the WoT lines,  a number of different approaches (in terms of, e.g., supporting middleware \cite{Dustdar15,Avatar15} and programming approaches \cite{Beal15,swarmlets}) are being proposed to support the engineering of IoT systems and applications. However, a common unifying approach supporting their design and development, grounded on a common set of abstractions, models, and methodologies, is still missing. This undermines the possibility of promoting a systematic and disciplined approach towards the development of complex IoT systems, and thus limits unfolding the full potentials of the IoT vision.

Against this background, this paper attempts at framing some key general-purpose issues related to the engineering of complex IoT systems and applications, by synthesizing the common characteristics of existing proposals and application scenarios. Such common characteristics are then used to identify the key software engineering abstractions around which the process of designing and developing IoT systems and applications could revolve, and via which to organize a set of guidelines towards the definition of a general methodology for engineering IoT systems.
 
To this end, the paper introduces a specific case study scenario, yet representative of a larger class of IoT scenarios. In particular, we consider the case of a big hotel with conference center. We assume the hotel infrastructures (e.g., lightening, heating, etc.)  and its facilities (guest rooms, conference rooms, and their associated appliances and objects) are densely enriched with connected sensors and actuators. In such scenario, as it will be discussed in the following, different actors can contribute to set up IoT services to support both the hotel management and the activities of its guests, but this requires the development of such services and of the overall IoT software infrastructure to be grounded on a sound methodological approach.    
 
The remainder of this paper is organized as follows. Section \ref{background} introduces some background concepts about IoT and WoT. On this basis, Section \ref{abstractions} introduces some key concepts and abstractions central to the engineering of IoT systems. Section \ref{methodology} sketches some guidelines for a methodology for the design and development of IoT systems. Section \ref{related} discusses related software engineering approaches, and Section \ref{conclusions} conclude.
 
\section{Background}
\label{background}

Most of the current approaches to the IoT envision a future in which millions of ICT sensors, actuators, and services, will be called to operate in an orchestrated way, as well as with the active contribution of the sensing, actuating, and reasoning capabilities of humans. This section overview some general background concepts that characterize most current visions of the IoT, as well as some general engineering issues.

\subsection{Things}

The ``things'' in the IoT vision may encompass a very large number of physical objects, and also include physical places and persons.

Physical objects can be made trackable and controllable via wireless by simply integrating with proper low-cost electronic devices. At the lower end of the spectrum, RFID tags or bluetooth beacons, based on low-cost and short-range communication protocols, can be attached to any kind of objects to enable tracking their positions, and possibly to associate some digital information with them. In between, one can think at more advanced RFID and beacons that integrates some environmental or motion sensors (i.e., accelerometers), to detect the present and the past activities associated with such objects. In addition, one can think at making objects actuable --  enabling the remote control of their configuration/status via proper digitally-controller actuators -- and possibly autonomous -- delegating them of autonomously direct their activities. 

To exemplify, in the hotel scenario: attach RFID tags to objects in rooms, such as to a remote control in order to detect its presence and location in the room; integrate some kind of Arduino-link controller to a roll-up board in the conference room, in order to enable controlling via, e.g., a mobile phone its rolling-unrolling; have the window obscuring systems autonomously regulate lightening conditions depending on the kind of activities detected in the conference room. In this perspective, autonomous robots (or robotified objects \cite{levolved}) can be somehow considered the highest end of the spectrum in the world of smart ``things''. 

For places, similarly to objects, one can attach simple RFID tags or beacons to places in the hotel, e.g., to detect the proximity of people to that places, or to sense specific environmental conditions in that places (at least for sensor-enriched tags and becons). Also, one can think at making remotely actuable parts of the environment, such as windows doors and walls. In the hotel scenario, one can think at rooms in which all the entities affecting environmental conditions (e.g., thermostats, windows, etc.) can be remotely controlled. In the hotel conference center, one can think at actuable walls that can dynamically change the shape and dimensions of meeting rooms depending on needs \cite{slothbot}.

Concerning persons, they can be perceived at first-class entities of the overall IoT vision \cite{Atzori14_2}. Simply for the fact of having a mobile phone, they can be somewhat sensed in their activities and positions, and they can be asked to act in the environment or supply sensing \cite{Hachem14}. In the hotel scenario, one may think continuously detecting the position and activities of people, in order to get ready to manage any possible emergency situation in the most efficient way from the viewpoint of people safety. Beside that, people in an environment also have the totally different role of exploiting the overall infrastructure of digitally enriched objects and places to get the best from their living in that environment, there include interacting with the physical environment and enriching their social experience in that environment, as described in Subsection \ref{services}. 

\subsection{Software Infrastructures}

To make IoT systems usable and capable of serving purposes, there is need of software infrastructures (that, an IoT middleware'' \cite{Teix11,surmid11}) capable both of supporting the ``gluing'' of different things and of providing some means for stakeholders and users to access the IoT system and take advantage of its functionalities.

Concerning the ``glue'', this involves a variety of technical issues:

\begin{itemize}

\item \emph{Interoperability}. To enable a variety of very heterogeneous things to  interact with each other, a set of shared tele-communication protocols and data representation schemes must be put in place \cite{context}, other than means to identify things \cite{identities}.  The study of these issues dates to the very early stages of IoT researches, a number of different proposals exists, and the way towards assessed standards in well paved.

\item \emph{Semantics}. Beyond mere interoperability, a common semantics for concepts must be defined to enable cooperation and integration of things \cite{semantics}. Also for this issue, a number of proposals grounded on standard Web technologies Web and defining ontologies and schemas specifically suited for the physically and socially embedded nature of the IoT exists \cite{context}.

\item \emph{Discovery, Group Formation, and Coordination}. Most of the functions provided in the context of IoT system derives from the orchestrated access and exploitation of a variety of things, possibly involving a variety of users and stakeholders. For instance, in the hotel scenario, the need to configure a conference rooms for slide presentation requires involving the beam projector, the lightening system, and consider the involvement of the conference organizers and of the speakers. This requires means to discovery and establish relations between things, and between things and humans \cite{Atzori14_2}, and coordinating their activities also accounting for their social relations \cite{Atzori14}.

\item \emph{Context-awareness and self-adaptation}. The large number of connected things that will define future IoT scenarios, and their inherent ephemerality, unreliability, and mobility (e.g., things such as chairs or flipboard in the hotel conference centre can come and go, can get moved around, and can be placed sometimes in corners with not wireless connections) makes it impossible to anticipate which things will be available and for how long during their exploitation. This requires that the mechanisms for discovery, group formation, and coordination are capable of dynamically adapting to the general context in which they act \cite{YeDM12}, and that mechanisms are provided for such groups to dynamically self-adapt on-the-fly, or possibly even self-organize in a context-aware way \cite{selforgcoord}.

\end{itemize}

Concerning the ``access'' to the functionalities and capabilities of individual things by users, as well as the orchestration of the functionalities of groups of things, the scene is currently dominate by the so called ``Web of Things'' (WoT) vision \cite{wot15}. The idea is to expose services and functionalities of individual things in terms of REST services, each associated to a URI and to be simply accessed via http GET and POST calls. This makes also possible to rely on assessed web technologies as far as discovery of things and provisioning of coordinated group services are concerned. 

In the past few years, a variety of proposals for middleware infrastructures to support the provisioning of IoT services and applications have appeared \cite{Dustdar15,Fortino12,Beal15,swarmlets}. Beside the specificities of the different proposals, most of them rely on: some basic infrastructure to support the WoT approach (i.e., to expose things in terms of simple services); some means of supporting, in according to some specific coordination model, the discovery of things (and of their associated services), and the coordinated activities of groups of things; some solutions to make services and applications capable of self-adapting and self-organizing in a context-aware and unsupervised way.

\subsection{Services and Applications}
\label{services}

With the term ``IoT System'' we generally refer to the overall set of IoT devices and to the associated middleware devoted to manage their networking and interactions. Logically above an IoT system, specific software can be deployed to orchestrate the activities of the system so as to provide:

\begin{itemize}

\item A number of specific \emph{services}, that is means to enable stakeholders and users to access and exploit individual things and direct/activate their sensing/actuating capabilities, but also coordinated services that access groups of things and coordinate their sensing/actuating capabilities. For instance, in a conference room of the hotel, other than to services to access and control individual appliances, one can think at providing a coordinated service that, by accessing and directing the lightening system, the light sensors, and the windows obscuring system, can modify the overall situation of the room from ``presentation state'' to ''discussion state'' and viceversa.

\item A number of more general-purpose \emph{applications}, intended as more comprehensive software systems intended to both regulate the overall functioning of an IoT system (or of some of its parts), so as to ensure specific overall behaviour of the system, as well as to provide an harmonized set of service to access the system and (possibly) its configuration. In the hotel scenario, one can think at applications to control the overall heating systems and lightening systems, and giving to hotel clerks the access to services to change the configuration of the associated parameter.

\end{itemize}

Clearly, depending on the specific scenario, one can think at IoT systems in which services may exist only confined within the context of some general application, but also at scenarios in which there are services that can be deployed as stand-alone software.

Given a specific IoT system, powered by a specific software infrastructure, a large variety of systems and applications can be designed and deployed over it, depending on the specific purpose it is intended to serve, and the specific classes of user that will somewhat exploit its services and applications. To make sure that the design and development of IoT services and applications fulfill expectations in a dependable way, though, a disciplined and rigorous approach to software analysis, design, and development, is necessary. This is definitely a general concern in software engineering, which is made even more relevant by the pervasive nature of IoT systems and by their primary role in supporting our everyday activities.

\section{Key Software Engineering Concepts and Abstractions}
\label{abstractions}

\begin{figure}
 \includegraphics[width=1\columnwidth]{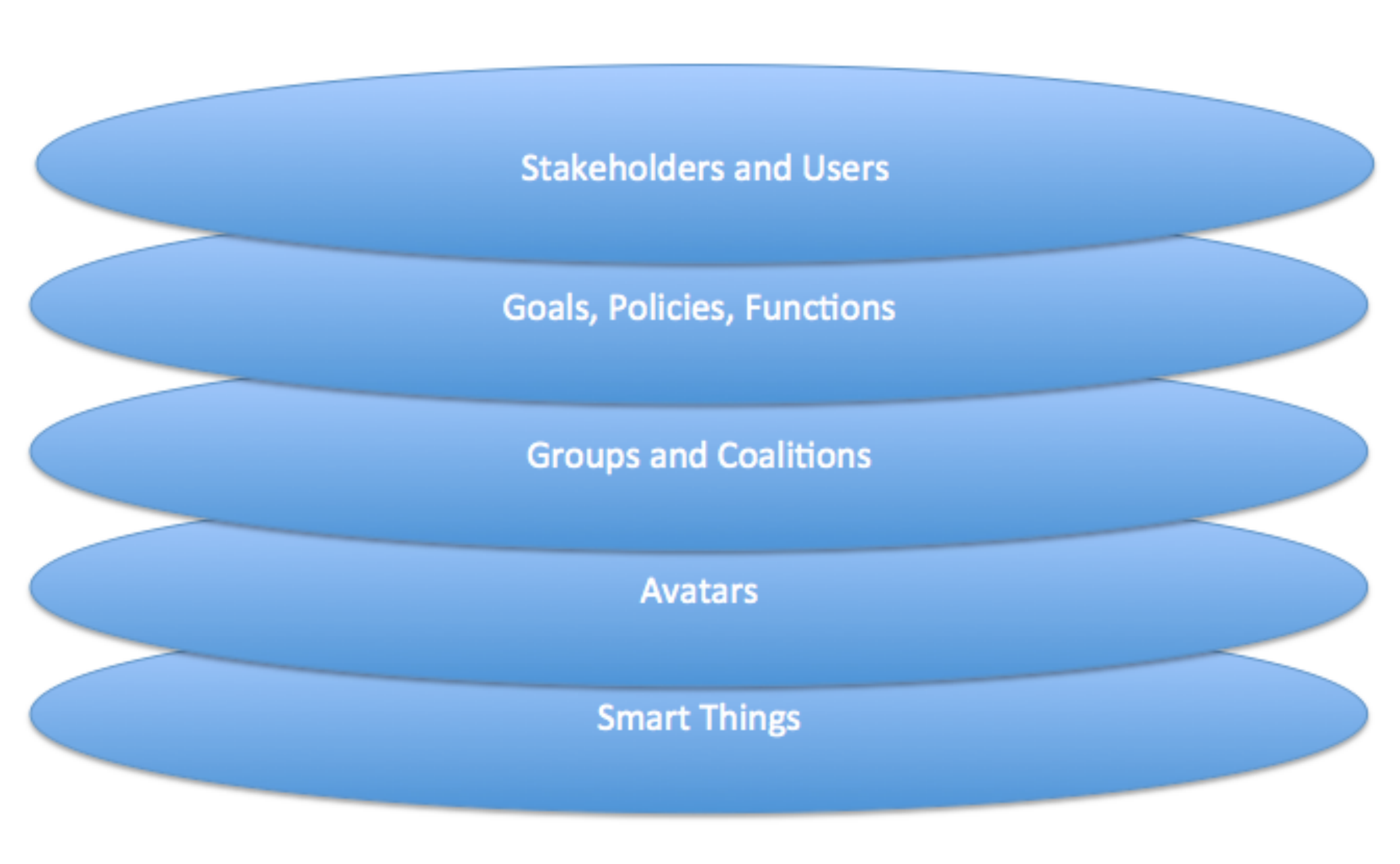}
  \caption{The stack with the key concepts and abstractions for IoT systems engineering.}
  \label{img:stack}
\end{figure}

Based on the above overview of IoT issues, in this section we try to synthesize the central concepts and abstractions around which the development of IoT systems (spanning analysis, design and implementation) should be centered. Figure \ref{img:stack} graphically frames such concepts in a logical stack.

\subsection{Stakeholders and Users}

The primary activities in the analysis of a system-to-be concern identifying the stakeholders and the user of the systems. That is, those persons/organizations who will  own, manage, and/or use the system and its functionalities, and from which requirements should be elicited, aka the ``actors''. 

In the case of IoT systems, there are three main classes of ``actors'' that can be identified: 

\begin{itemize}

\item \emph{Global Managers}: These are typically the owners of an overall IoT system and infrastructure, or at least persons empowered to exert control over the configuration, structure, and overall functioning of its applications and services. In the hotel scenario, the global manager corresponds to the hotel management, e.g., the system manager devoted to control the overall IoT system of the hotel according to the directives of the hotel management. The global manager can establish the policies according to which to run the overall infrastructure and its applications, e.g., for deciding heating levels or for surveillance strategies. 

\item \emph{Local Managers}: These are typically owners (whether permanently or on a temporary basis) of a limited portion of the IoT system, or their delegates, and are empowered to enforce local control for that portion of the infrastructure. In the hotel scenario, these could correspond to hotel guests, which can be empowered to control the IoT system in their room, and tune the local parameters and exploit its services according to their own specific needs. Or they can be the organizers of a conference in charge of managing and configuring the services of the rented conference rooms.

\item \emph{Users}: These are typically persons or group that have limited access to the overall configuration of the IoT applications and services, i.e., cannot impose policies on them, but are nevertheless entitled to exploit its services. In the hotel scenario, these can include conference delegates that are authorized to access the conference facilities (e.g., uploading presentations in the projector and regulating the microphone and lightening in the meeting room), but are not entitles to modiy the basic configuration of the infrastructure.

\end{itemize}

The three identified classes of actors are of a very general nature, beside the hotel scenario. For example, considering a scenario of energy management in a smart city, they could be made corresponding to, respectively: city managers, house/shop owners, private citizens and tourists. In the area of urban mobility, they could be made corresponding to, respectively: mobility managers, parking owners or car sharing companies, private drivers.

\subsection{Functionalities}

The development of IoT applications and services cannot simply reduce to understand the functionalities that objects or group of objects has to provide, but has to account for a more comprehensive approach, accounting for the following facts: 

\begin{itemize}

\item Beside things provided with basic sensing/actuating functionalities, one should consider the presence of smarter things that can be activated to perform in autonomy some long-term activities associated with their nature and with their role in the socio/physical environment in which they situates. These can range from simply cleaning robots to more sophisticated autonomous personal assistants \cite{levolved}.

\item IoT applications are not simply concerned with providing a suite of coordinated functionalities, but they should also globally regulate the activities of the IoT systems on a continuous basis, according to the policies established by its stakeholders and to their objectives.  

\end{itemize}

As a consequence, developing IoT services and applications, other than defining and implementing functionalities, most often implies defining \emph{policies} and \emph{goals} associated to services and applications \cite{van2001goal}. In general terms, policies and goals represents desirable ``state of the affairs'' to strive for. In the context of an IoT system, policies and goals represents specific configurations of the overall socio-cyber-physical system (or of portion of it) that IoT applications and services are in charge of eventually producing and/or maintaining, respectively. 

In this perspective, the overall functionalities to be defined by IoT applications and services can be framed as follows: 

\begin{itemize}

\item \emph{Policies}. These expresses desirable permanent configurations or states of functioning of an overall IoT system (global policies) or portions of it (local policies), and have the aims of regulating the overall underlying IoT system. In the hotel scenarios, global policies can be defined, e.g., to specify the maximum occupancy levels in each room and have this monitored by local cameras in order to invite people to move in different rooms whenever needed. Policies are meant to be alway active and actively enforced. Although, from the software engineering viewpoint, the focus is mostly on application-level policies, policies can also account for the proper configuration of the underlying hardware and network infrastructures, in line with a software-defined networking perspective \cite{cassowary}. The definition of global and local policies is generally in charge of the global managers, although local managers can be also entitled to enforce temporary local policies on local portions of the system (provided they do not contrast with the ones imposed by the global managers).

\item \emph{Goals}. These express desirable situations or state of the affairs that, in specific cases, can/should be activated. The activation of a goal may rely on specific pre-conditions (i.e., the occurrence of specific events or the recognition of some specific configurations in the IoT system) or may also be specifically activated upon user action (e.g., if the activation of a goal is invokable as a service). The typical post-condition (deactivating the pursuing of a goal) is the achievement of the goal itself, although one can also consider post-conditions for terminating the goal earlier than its achievement. In the hotel scenario, the clearer example could be that of activating an evacuation procedure upon detection of fire by some sensors (pre-condition), whose goal (and post-condition) is to achieve a quick evacuation of all people inside the building. To this end, the activation of a goal can trigger the activities of digital signages and controllable doors in order to rationally guide people towards the exits. As for policies, the definition of global and local global is generally in charge of global, and sometimes of local, managers. 

\item \emph{Functions}, define the sensing/computing/actuating capabilities of individual things or of group of things, or the specific resources, that are to be made available to managers and users in the context of specific IoT application and services. Functions are typically made accessible in the form of services, and can sometime involve the orchestrated access to the functions of a multitude of individual things. In the hotel scenario, one can think at the individual functionalities of the appliances in a conference room (e.g., open/close a curtain, display slide / change slide in a projector), as well as more complex functionalities that can be achieved by orchestrating things (e.g., set up room for presentation by closing all curtains and switching off all lights). Functions and the associated services are typically defined by global and possibly local managers, but are exploited also by the everyday users of the IoT systems (e.g., the hotel guests and the conference attendees).

\end{itemize}

\begin{figure}
 \includegraphics[width=0.8\columnwidth]{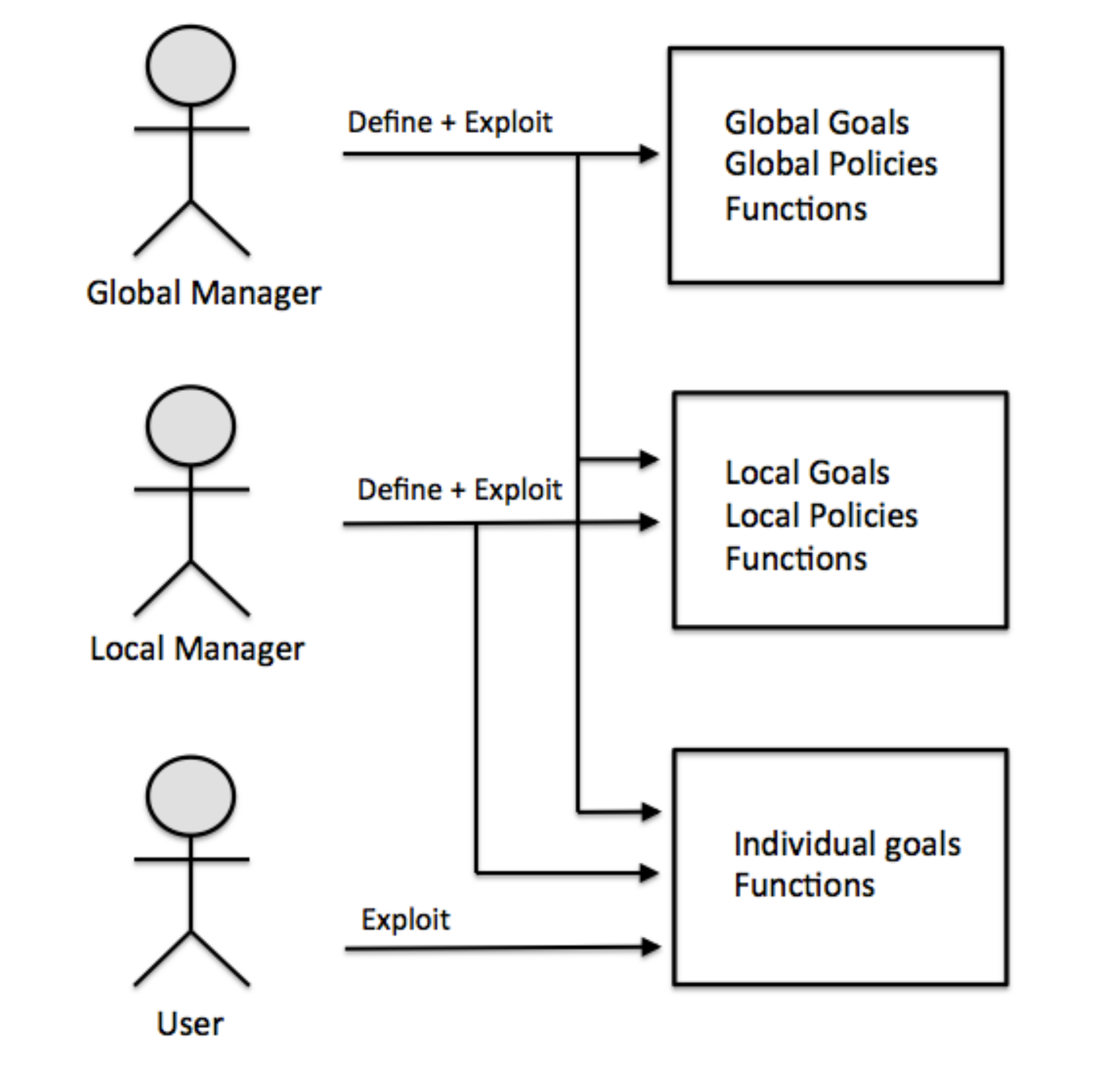}
  \caption{The actors and the functionalities of IoT systems.}
  \label{img:usecases}
\end{figure}

Figure \ref{img:usecases} shows the different roles of IoT actors in defining and exploiting the functionalities of IoT systems.

\subsection{Avatars and Coalitions}

The ``things'' in an IoT system can correspond to a variety of different objects and ICT devices, other than to places and humans, each relying on a pletora of different technologies and capabilities. Accordingly, from both the gluing software infrastructure and the software engineering viewpoints, it is necessary to define higher-level abstractions to practically and conceptually handle the development of application and services, and to harmonically exploit all the components of the IoT system. 

Most of the proposal for programming models and middleware acknowledge this need, by means of a software layer in which individual things are virtualized in some sort of software abstraction \cite{wot15}. The web of thing perspective suggests abstracting things and their functionalities in terms of generic resources, to be accessed via RESTful calls \cite{wot09,wottuto}, possibly associating specific external software HTTP ``gateways'' to individual things, whenever they cannot directly support HTTP interfacing \cite{gateways}. Other approaches suggest adopting a more stanrdard SOA or object-oriented approach \cite{iotlayered}. Also, some proposals consider associating autonomous software agents to individual things \cite{AALagents}, which we think well suits the fact that goals to be pursued in autonomy may be associated to things.  

In addition, as already stated, some ``things'' make no sense as individual entities as far as the provisioning of specific services and applications is concerned, and are to be considered part of a group and be capable of providing their services as a coordinated group. This applies both to the cases in which a multitude of equivalent devices must be collectively exploited, in order to abstract from the presence of the individuals and rather relying on the capabilities of the group \cite{Beal15,Bic12}, and to the cases in which the functionalities of the group complement with each other and needs to be orchestrated \cite{Teix11,AALagents}.

\begin{figure}
 \includegraphics[width=1\columnwidth]{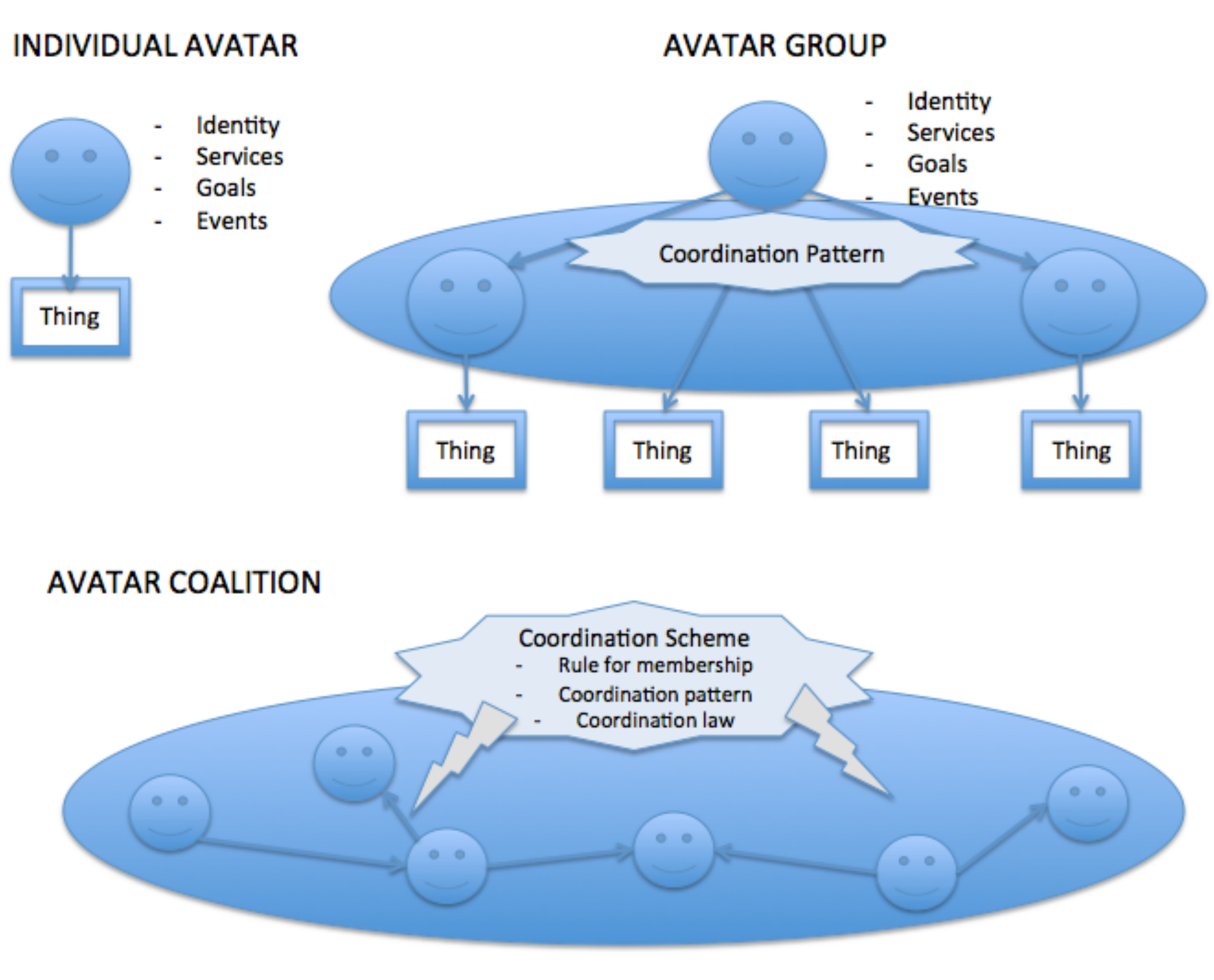}
  \caption{Avatars, groups, and coalitions.}
  \label{img:avatars}
\end{figure}

With these considerations in mind, and in an effort from synthesizing from a variety of different proposals, we suggest the following unifying abstractions (See Figure \ref{img:avatars}):

\begin{itemize}

\item \emph{Avatars}. Borrowing the term from \cite{Avatar15}, we define an avatar as the general abstraction both for individual things and for group of things that logically define a unique concept. Avatars abstract away form the specific physical/social/technological characteristics of the things their represent, as well as (for the case of groups) from the specific coordination techniques that are used to coordinate individual things within the group, and are defined by means of:

\begin{itemize}

\item \emph{Identity}. An avatar has a unique identity and is addressable. An avatar enclosing a group of avatar does not necessarily hides the identities of inner avatars, but it has its own identity.

\item \emph{Services}. Services represents access point for exploiting the peculiar capabilities of avatars. That is, depending on the kinds of things and functionalities it abstract: triggering and directing the sensing/computing/actuating capabilities, or accessing some managed resources.

\item \emph{Goals}. Goals, in the sense of desired state of the affairs, can be associated to Avatars. Goals have may a pre-condition for autonomous activation, or may be explictly activated by a user or by another avatar.

\item \emph{Events}. Events represent specific state of the affairs that can be detected by an avatar, and that may be of interests to other avatars or to users. Other avatars or users can subscribe to events of interest.

\end{itemize}

Clearly, for group of avatars, an internal \emph{Coordination Pattern} must be defined to orchestrate the activities/functionalities of the objects (or of the other avatars) it includes. In general terms, a coordination patterns defines the internal workflow of activities among the composing objects and avatars, and the constrains/conditions they are subjected to. Coordination patterns may also account for contextual information and thus making the activities of the group of avatar adaptive to changes in the context. 

The above abstraction is perfectly in line, and account for all the required characteristics of the systems we already cited. The idea is not fully in line with that of RESTful Web-based approaches, because of the stateful concepts of goals and events. However, it is to be said that most WoT approach recognize the need to somehow incorporate similar concepts even within RESTful approaches \cite{wottuto}, because they are necessary to suit the dynamic and contextual nature of IoT systems and applications. 

\item \emph{Coalitions}. Borrowing the term from the area of multiagent systems \cite{coalitions}, here we define a coalition as a group of avatars that coordinate each other's activities in order to reach specific goals, or enact specific policies. Accordingly, coalitions may be of a temporary or permanent nature. Unlike avatars, coalitions does not necessarily have an identity, and does not necessarily provide services. To define and bring a coalition in action, the abstraction of coalition must be defined (at least) in terms of a \emph{coordination scheme} that should include:

\begin{itemize}

\item \emph{Rules for membership}. This specify the specific conditions upon which an avatar should/could enter a coalitions. From the viewpoint of individual avatars, the act of entering a coalition can be represented by the activation of a specific goal based on pre-conditions that correspond to the rules for membership.

\item \emph{Coordination pattern}. This define the pattern (interaction scheme and workflow of activities) by which the members of the coalition have to interact. The coordination pattern may include an explicit representation of the goal by which the coalition has been activated. However, such goal can also be implicit in the definition of the pattern and of the workflow. 

\item \emph{Coordination law}. This expresses constraints that must be enforced in the way the avatars involved in the coalition should act and interact. 

\end{itemize}
\end{itemize}

In addition, one can consider the possibility to subscribe to events occurring within the coalition.

The view of avatar coalitions can be of use to realize policies, or to aggregate groups of avatar based on similarity, so as to make them work collectively without forcing them to specific orchestration. This is coherent with the idea of aggregate programming in sensor networks \cite{Bic12} and in spatial computing systems \cite{Beal15}, to realize nature-inspired coordination schemes \cite{Fer13,Zambonelli15}, to enable the dynamic formation of groups focused on short-term goals \cite{Jennings14,SassiZ14}, or to define ensembles of services based on specific attributes \cite{scel}.

\section{Towards a Software Engineering Methodology}
\label{methodology}
As is the case with any new software paradigm and technological scenario, the successful and widespread deployment of complex software applications and services for the IoT requires not only the identification of the proper software engineering abstraction -- which we have attempted in the previous section -- but also the identification of an appropriate software engineering \emph{methodology} -- i.e., a set of guidelines revolving around such abstractions and facilitating engineers in developing such systems in a robust, reliable, and repeatable fashion. 

\begin{figure}
 \includegraphics[width=1\columnwidth]{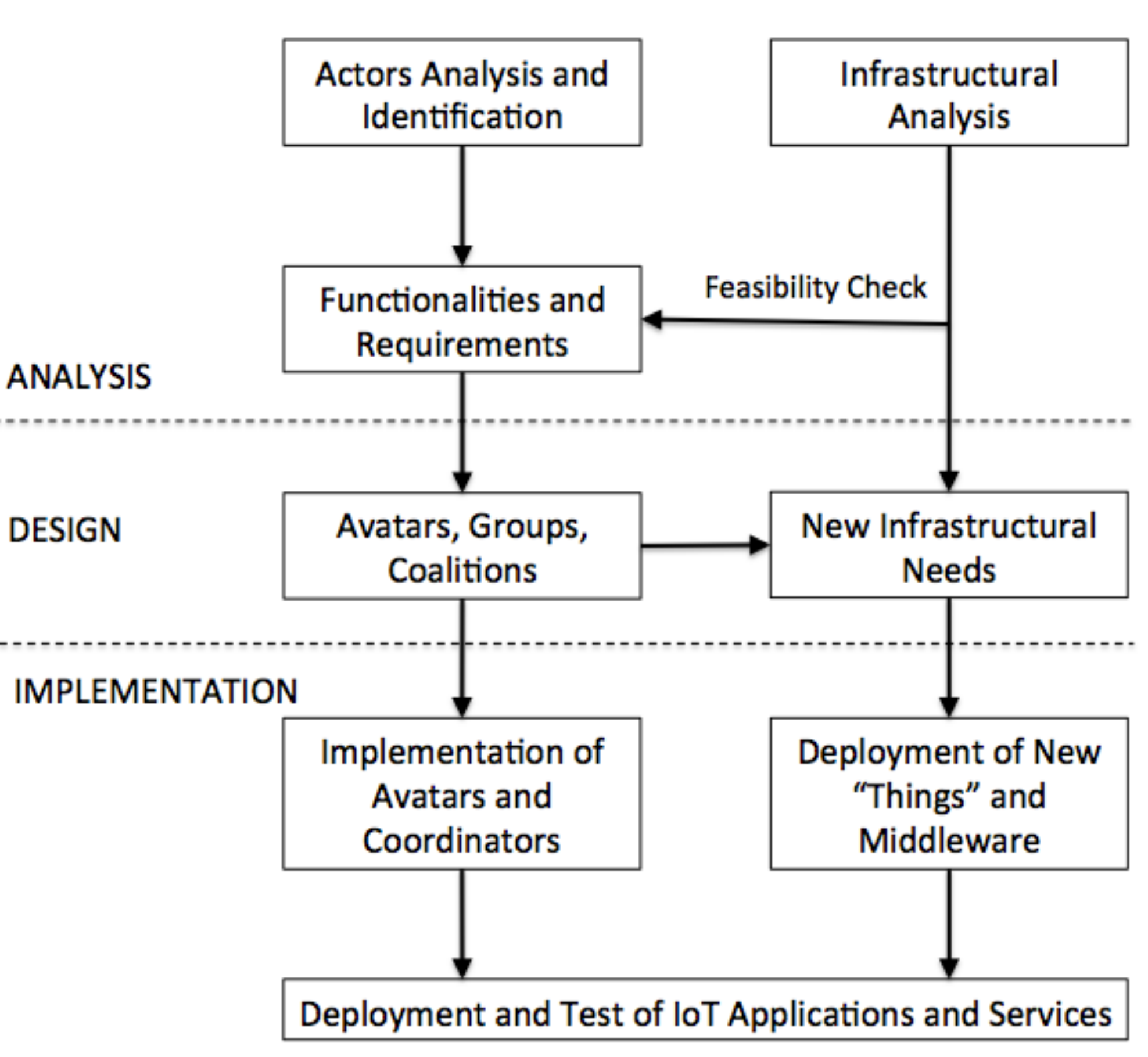}
  \caption{Overview of a general IoT methodology.}
  \label{img:methodology}
\end{figure}

The definition of a complete software engineering methodology is a very complex issue, and should  rely on a large body of real-word experiences. Also, it and should be accompained by a proper set of models and tools via which to represent and produce -- in accord with the basic abstractions of the paradigm -- the conceptual and software artifacts that will eventually lead to the final product. Nevertheless, based on the current state of documented experiences, and given the analysis of the central abstractions provided above, it is possible to sketch some general guidelines and identify the different steps of the software process (See Figure \ref{img:methodology}).  

For the \emph{analysis phase}, the envisioned activities include:

\begin{itemize}

\item \emph{Actors analysis and identification}. Beside global managers, which are the primary stakeholders, this activity implies (in cooperation with the global managers) identifying the characteristics of the users of the systems, as well as of those users which can be entitled to play the role of local managers.

\item \emph{Infrastructure analysis}. Increasingly in the future, new IoT applications and services will have to be deployed over an existing IoT hardware infrastructure and possibly of some associated middleware. Consequently, analyzing the characteristics and limitations of such infrastructure will be a necessary pre-requisite for the development of software over it.

\item \emph{Functionality and requirements}. This activities implies identifying goals, policies, functions, and define whether goals and policies are of a local or a global nature. This activity of requirements elicitation can involve mostly global managers, which (beside having to define global goals and global polities) are entitle to decide which services to provide to end users and which local goals and policies can be set up by local managers. As in modern participatory approaches to system design \cite{designthinking}, the possible involvement in this activity of potential end users and local managers could provide a more complete identification of needs and expectation. The identification of requirements should explicitly consider the characteristics of the infrastructure, in that some of the requirements may not be feasible on it, and may require the later integration of some new features of the infrastructure.

\end{itemize}

For the \emph{design phase}, the envisioned activities include:

\begin{itemize}

\item \emph{Design of Avatars, Groups and Coalitions}. This must be defined with the goal of producing an overall design that is capable to realize the necessary goals. policies, and functions, in accord to the requirements. We emphasize that the design of coalitions does not imply designing new avatars, but primarily at designing the coordination scheme that will define and rule the activities of the coalition.

\item \emph{Identification of new infrastructural needs}. To understand what new devices or middleware services that must be integrated in the infrastructure in order to realize the needed functionalities in accord with the requirements. 

\end{itemize}

For the \emph{design phase}, the envisioned activities include:

\begin{itemize}

\item \emph{Implementation of Avatars and Coordinators}. This clearly depend on the adopted middleware infrastructure and programming model. However, the abstraction of avatars and the concept of coordinators (intended as a software artifact to support a specific coordination scheme via which to realize coalitions) are of a very  general nature, and can apply to a large-variety of actual IoT middleware and programming models.

\item \emph{Deployment of new things and new middleware characteristics}. This will take place accordingly to the new infrastructural needs identified in the design phase.

\end{itemize}

Clearly, the above proposed methodology is not complete, and requires the definition of models and tools via which to represent the different conceptual and software artifacts that are produced in the various steps (e.g., how we model an IoT requirement? Via which formal or semi-formal language can we represent the characteristics of an avatar or of a group? How can we turn such abstract representation into an implementation?). This is left for future work.

\section{Related Work}
\label{related}

In the past few years, research in the area of IoT has exploded. Nevertheless, a few research work has explicitly attacked the problem of defining new software engineering approaches specifically conceived for the IoT. 

Some proposals for development frameworks for the IoT or for the WoT (whether middleware architecture \cite{Avatar15} or programming models \cite{Beal15,swarmlets}), are also accompanied by guidelines towards the development of applications. However, such guidelines are not grounded on general abstractions and haven't a general applicability beside the specific framework in which they are conceived. Similar considerations apply to the area of smart cities and urban computing \cite{Zambonelli12}, where middleware and programming approaches are being proposed -- mostly of a special-purpose nature and focussed on specific application scenarios such as participatory sensing \cite{Hachem14,Harnie14} or mobility management \cite{SassiZ14} -- but without accounting for the issue of defining general design and development methodologies. 

Agent-oriented software engineering research is strictly related to IoT engineering \cite{ZamO04}. Indeed, AOSE tackles the problem of engineering large-scale systems, goal-oriented entities, possibly including robots \cite{Scerri05} and humans \cite{Jennings14} with conflicting goals and a multitude of stakeholders. This is somehow related to the IoT problems of accommodating services and a multitude of goals \cite{AALagents}. Indeed, the idea of goal-oriented groups we have introduce somehow borrow from the agent-oriented software engineering area. However, IoT requires the introduction of specific concepts and abstractions that AOSE, in general terms, do not address.

General approaches for the engineering of self-organizing computing systems have been proposed \cite{Par97,ZamV11,selforgcoord,Zambonelli15,Fer13}. There, the key issue is to engineering complex distributed behaviours in large-scale systems lacking centralized control. These two characteristics are mostly shared by IoT systems, and indeed the problems of enabling self-organization of specific behaviors have been outlined in the previous sections. 

Mainstream software engineering researches have recently put great attention to the problem of promoting self-adaptive features in software \cite{roadmap}, to attack the problem of increased dynamically and impredictability of operational environments. Such dynamics also affects IoT systems, in which the problem of ensuring continuity in functionalities requires the embedding of close control loops (along similar lines of those promoted in self-adaptive systems researches) to continuously monitor the activities of the system and its environment, and eventually plan corrective actions.

\section{Conclusions and Future Work}
\label{conclusions}

Despite the large number of research works that, at many levels, attempt at attacking specific problems related to the design and development of IoT applications and services, a general principled software engineering approach is still missing. This paper, by having to framed the key conceptual abstractions revolving about the IoT universe, and by having sketched a methodology centered around these concept, can represent a first small step towards a general discipline for engineering IoT systems and applications.

As IoT technologies mature, and real-world experiences accumulate, more research in the area of software engineering for IoT systems and applications will be needed, possibly exploiting contaminations with the relevant areas of agent-oriented software engineering \cite{ZamO04} and software engineering for self-adaptive and self-organizing systems \cite{Fer13,roadmap}.

\vspace{3mm}

\bibliographystyle{plain}
\bibliography{IOT}

\begin{thebibliography}{10}

\bibitem{levolved}
Harshit Agrawal, Sang-won Leigh, and Pattie Maes.
\newblock L'evolved: Autonomous and ubiquitous utilities as smart agents.
\newblock In {\em Proceedings of the 2015 ACM International Joint Conference on
  Pervasive and Ubiquitous Computing}, UbiComp '15, pages 487--491, New York,
  NY, USA, 2015. ACM.

\bibitem{Atzori14}
Luigi Atzori, Davide Carboni, and Antonio Iera.
\newblock Smart things in the social loop: Paradigms, technologies, and
  potentials.
\newblock {\em Ad Hoc Networks}, 18:121--132, 2014.

\bibitem{Atzori14_2}
Luigi. Atzori, A.~Iera, and G.~Morabito.
\newblock From ``smart objects'' to ``social objects'': The next evolutionary
  step of the {I}nternet of {T}hings.
\newblock {\em IEEE Communications Magazine}, 52(1):97--105, January 2014.

\bibitem{Atzori10}
Luigi Atzori, Antonio Iera, and Giacomo Morabito.
\newblock The internet of things: {A} survey.
\newblock {\em Computer Networks}, 54(15):2787--2805, 2010.

\bibitem{surmid11}
Soma Bandyopadhyay, Munmun Sengupta, Souvik Maiti, and Subhajit Dutta.
\newblock A survey of middleware for internet of things.
\newblock In Abdulkadir Özcan, Jan Zizka, and Dhinaharan Nagamalai, editors,
  {\em Recent Trends in Wireless and Mobile Networks}, volume 162 of {\em
  Communications in Computer and Information Science}, pages 288--296. Springer
  Berlin Heidelberg, 2011.

\bibitem{semantics}
Payam Barnaghi, Wei Wang, Cory Henson, and Kerry Taylor.
\newblock Semantics for the internet of things: early progress and back to the
  future.
\newblock {\em International Journal on Semantic Web and Information Systems
  (IJSWIS)}, 8(1):1--21, 2012.

\bibitem{Beal15}
Jacob Beal, Danilo Pianini, and Mirko Viroli.
\newblock Aggregate programming for the internet of things.
\newblock {\em {IEEE} Computer}, 48(9):22--30, 2015.

\bibitem{Bic12}
N.~Bicocchi, M.~Mamei, and F.~Zambonelli.
\newblock Self-organizing virtual macro sensors.
\newblock {\em ACM Transaction on Autonomous Adaptive Systems}, 7(1), 2012.

\bibitem{gateways}
G{\'e}r\^{o}me Bovet and Jean Hennebert.
\newblock Offering web-of-things connectivity to building networks.
\newblock In {\em Proceedings of the 2013 ACM Conference on Pervasive and
  Ubiquitous Computing Adjunct Publication}, UbiComp '13 Adjunct, pages
  1555--1564, New York, NY, USA, 2013. ACM.

\bibitem{coalitions}
Yongcan Cao, Wenwu Yu, Wei Ren, and Guanrong Chen.
\newblock An overview of recent progress in the study of distributed
  multi-agent coordination.
\newblock {\em Industrial Informatics, IEEE Transactions on}, 9(1):427--438,
  2013.

\bibitem{roadmap}
B.~H.~C. Cheng and al.
\newblock Software engineering for self-adaptive systems: A research roadmap.
\newblock In {\em Software Engineering for Self-Adaptive Systems}, volume 5525
  of {\em Lecture Notes in Computer Science}, pages 1--26. Springer, 2009.

\bibitem{Con12}
M.~Conti, S.~Das, C.~Bisdikian, M.~Kumar, L.~Ni, A.~Passarella, G.~Roussos,
  G.~Troster, G.~Tsudik, and F.~Zambonelli.
\newblock Looking ahead in pervasive computing: Challenges and opportunities in
  the era of cyber-physical convergence.
\newblock {\em Pervasive and Mobile Computing}, 8(1):2 -- 21, 2012.

\bibitem{scel}
Rocco {De Nicola}, Diego Latella, Alberto Lluch{-}Lafuente, Michele Loreti,
  Andrea Margheri, Mieke Massink, Andrea Morichetta, Rosario Pugliese,
  Francesco Tiezzi, and Andrea Vandin.
\newblock The {SCEL} language: Design, implementation, verification.
\newblock In {\em Software Engineering for Collective Autonomic Systems - The
  {ASCENS} Approach}, pages 3--71. 2015.

\bibitem{Fer13}
J.L Fernandez-Marquez, G.~Di~Marzo Serugendo, S.~Montagna, M.~Viroli, and
  J.~Arcos.
\newblock Description and composition of bio-inspired design patterns: a
  complete overview.
\newblock {\em Natural Computing}, 12(1):43 -- 67, 2013.

\bibitem{Fortino12}
Giancarlo Fortino, Antonio Guerrieri, and Wilma Russo.
\newblock Agent-oriented smart objects development.
\newblock In {\em {IEEE} 16th International Conference on Computer Supported
  Cooperative Work in Design, {CSCWD} 2012, May 23-25, 2012, Wuhan, China},
  pages 907--912. IEEE, 2012.

\bibitem{gubbi2013}
Jayavardhana Gubbi, Rajkumar Buyya, Slaven Marusic, and Marimuthu Palaniswami.
\newblock Internet of things (iot): A vision, architectural elements, and
  future directions.
\newblock {\em Future Generation Computer Systems}, 29(7):1645--1660, 2013.

\bibitem{wot09}
Dominique Guinard and Vlad Trifa.
\newblock Towards the web of things: Web mashups for embedded devices.
\newblock In {\em Workshop on Mashups, Enterprise Mashups and Lightweight
  Composition on the Web (MEM 2009), in proceedings of WWW (International World
  Wide Web Conferences)}, Madrid, Spain, April 2009.

\bibitem{wottuto}
Dominique Guinard, Vlad Trifa, Friedemann Mattern, and Erik Wilde.
\newblock From the internet of things to the web of things: Resource-oriented
  architecture and best practices.
\newblock In Dieter Uckelmann, Mark Harrison, and Florian Michahelles, editors,
  {\em Architecting the Internet of Things}, pages 97--129. Springer Berlin
  Heidelberg, 2011.

\bibitem{Hachem14}
Sara Hachem, Animesh Pathak, and Val{\'{e}}rie Issarny.
\newblock Service-oriented middleware for large-scale mobile participatory
  sensing.
\newblock {\em Pervasive and Mobile Computing}, 10:66--82, 2014.

\bibitem{Harnie14}
Dries Harnie, Theo D'Hondt, Elisa~Gonzalez Boix, and Wolfgang De~Meuter.
\newblock Programming urban-area applications for mobility services.
\newblock {\em ACM Transactions on Autonomous and Adaptive Systems}, 9(2),
  2014.

\bibitem{wot15}
J.~Heuer, J.~Hund, and O.~Pfaff.
\newblock Toward the web of things: Applying web technologies to the physical
  world.
\newblock {\em Computer}, 48(5):34--42, May 2015.

\bibitem{Ian14}
Marco Iansiti and Karin Lakhani.
\newblock Digital ubiquity: How connections, sensors, and data, are
  revolutionizing business.
\newblock {\em Harvard Business Review}, 2014.

\bibitem{Jennings14}
N.~R. Jennings, L.~Moreau, D.~Nicholson, S.~Ramchurn, S.~Roberts, T.~Rodden,
  and A.~Rogers.
\newblock Human-agent collectives.
\newblock {\em Commun. ACM ACM}, 57(12):80--88, December 2014.

\bibitem{cassowary}
Pradeeban Kathiravelu, Leila Sharifi, and Lu\'{\i}s Veiga.
\newblock Cassowary: Middleware platform for context-aware smart buildings with
  software-defined sensor networks.
\newblock In {\em Proceedings of the 2Nd Workshop on Middleware for
  Context-Aware Applications in the IoT}, M4IoT 2015, pages 1--6, New York, NY,
  USA, 2015. ACM.

\bibitem{swarmlets}
E.~Latronico, E.A. Lee, M.~Lohstroh, C.~Shaver, A.~Wasicek, and M.~Weber.
\newblock A vision of swarmlets.
\newblock {\em Internet Computing, IEEE}, 19(2):20--28, Mar 2015.

\bibitem{Avatar15}
M.~Mrissa, L.~Medini, J.-P. Jamont, N.~Le~Sommer, and J.~Laplace.
\newblock An avatar architecture for the web of things.
\newblock {\em Internet Computing, IEEE}, 19(2):30--38, Mar 2015.

\bibitem{selforgcoord}
Andrea Omicini and Mirko Viroli.
\newblock Coordination models and languages: From parallel computing to
  self-organisation.
\newblock {\em The Knowledge Engineering Review}, 26(1):53--59, March 2011.

\bibitem{Par97}
Van Parunak.
\newblock Go to the ant: Engineering principles from natural multi-agent
  systems.
\newblock {\em Annals of Operations Research}, 75:69--101, 1997.

\bibitem{context}
C.~Perera, A.~Zaslavsky, P.~Christen, and D.~Georgakopoulos.
\newblock Context aware computing for the internet of things: A survey.
\newblock {\em Communications Surveys Tutorials, IEEE}, 16(1):414--454, First
  2014.

\bibitem{slothbot}
M.~Phillips.
\newblock The slothbot moving wall projects.

\bibitem{designthinking}
Hasso Plattner, Christoph Meinel, and Larry Leifer.
\newblock {\em Design Thinking: Understand--Improve--Apply}.
\newblock Springer Science \& Business Media, 2010.

\bibitem{Por14}
Michael Porter and James HeppelMann.
\newblock How smart connected products are transforming competition.
\newblock {\em Harvard Business Review}, 2014.

\bibitem{iotlayered}
C.~Sarkar, S.N.A.U. Nambi, R.V. Prasad, and A.~Rahim.
\newblock A scalable distributed architecture towards unifying iot
  applications.
\newblock In {\em Internet of Things (WF-IoT), 2014 IEEE World Forum on}, pages
  508--513, March 2014.

\bibitem{identities}
Amardeo~C Sarma and Jo{\~a}o Gir{\~a}o.
\newblock Identities in the future internet of things.
\newblock {\em Wireless personal communications}, 49(3):353--363, 2009.

\bibitem{SassiZ14}
Andrea Sassi and Franco Zambonelli.
\newblock Coordination infrastructures for future smart social mobility
  services.
\newblock {\em IEEE Intelligent Systems}, 29(5):78--82, 2014.

\bibitem{Scerri05}
Nathan Schurr, Janusz Marecki, Milind Tambe, and Paul Scerri.
\newblock Towards flexible coordination of human-agent teams.
\newblock {\em Multiagent and Grid Systems}, 1(1):3--16, 2005.

\bibitem{AALagents}
N.~Spanoudakis and P.~Moraitis.
\newblock Engineering ambient intelligence systems using agent technology.
\newblock {\em Intelligent Systems, IEEE}, 30(3):60--67, May 2015.

\bibitem{Teix11}
Thiago Teixeira, Sara Hachem, Val{\'e}rie Issarny, and Nikolaos Georgantas.
\newblock Service oriented middleware for the internet of things: A
  perspective.
\newblock In {\em Proceedings of the 4th European Conference on Towards a
  Service-based Internet}, ServiceWave'11, pages 220--229, Berlin, Heidelberg,
  2011. Springer-Verlag.

\bibitem{van2001goal}
Axel Van~Lamsweerde.
\newblock Goal-oriented requirements engineering: A guided tour.
\newblock In {\em Requirements Engineering, 2001. Proceedings. Fifth IEEE
  International Symposium on}, pages 249--262. IEEE, 2001.

\bibitem{Dustdar15}
Lina Yao, Q.Z. Sheng, and S.~Dustdar.
\newblock Web-based management of the internet of things.
\newblock {\em Internet Computing, IEEE}, 19(4):60--67, July 2015.

\bibitem{YeDM12}
Juan Ye, Simon Dobson, and Susan McKeever.
\newblock Situation identification techniques in pervasive computing: A review.
\newblock {\em Pervasive and Mobile Computing}, 8(1):36--66, 2012.

\bibitem{Zambonelli12}
Franco Zambonelli.
\newblock Toward sociotechnical urban superorganisms.
\newblock {\em IEEE Computer}, 45(8):76--78, 2012.

\bibitem{ZamO04}
Franco Zambonelli and Andrea Omicini.
\newblock Challenges and research directions in agent-oriented software
  engineering.
\newblock {\em Autonomous Agents and Multi-Agent Systems}, 9(3):253--283,
  November 2004.
\newblock Special Issue: Challenges for Agent-Based Computing.

\bibitem{Zambonelli15}
Franco Zambonelli, Andrea Omicini, Bernhard Anzengruber, Gabriella Castelli,
  Francesco L.~De Angelis, Giovanna Di~Marzo Serugendo, Simon Dobson, Jose~Luis
  Fernandez-Marquez, Alois Ferscha, Marco Mamei, Stefano Mariani, Ambra
  Molesini, Sara Montagna, Jussi Nieminen, Danilo Pianini, Matteo Risoldi,
  Alberto Rosi, Graeme Stevenson, Mirko Viroli, and Juan Ye.
\newblock Developing pervasive multi-agent systems with nature-inspired
  coordination.
\newblock {\em Pervasive and Mobile Computing}, 37, 2015.

\bibitem{ZamV11}
Franco Zambonelli and Mirko Viroli.
\newblock A survey on nature-inspired metaphors for pervasive service
  ecosystems.
\newblock {\em Journal of Pervasive Computing and Communications}, 7:186--204,
  2011.

\end{thebibliography}

\end{document}